\documentclass[prd,twocolumn,showpacs]{revtex4}

\usepackage{epsfig}
\usepackage{graphics}
\usepackage{amsmath}
\usepackage{bm}

\newcommand{\Eq}[1]{Eq.~(\ref{#1})}
\newcommand{\be}{\begin{equation}}
\newcommand{\ee}{\end{equation}}

\begin{document}
\title{Black holes in the TeVeS  theory of gravity and their thermodynamics }
\author{Eva Sagi}\email{eva.sagi@mail.huji.ac.il}
\author{Jacob D. Bekenstein}\email{bekenste@vms.huji.ac.il}
\affiliation{Racah Institute of Physics, Hebrew University of
Jerusalem, Jerusalem 91904, Israel\\}

\date{\today}

\begin{abstract}
TeVeS, a relativistic theory of gravity, was designed to provide a
basis for the modified Newtonian dynamics.  Since TeVeS differs from
general relativity (e.g., it has two metrics, an Einstein metric and
a physical metric), black hole solutions of it would be valuable for
a number of endeavors ranging from astrophysical modeling to
investigations into the interrelation between gravity and
thermodynamics.   Giannios has recently found a TeVeS analogue of
the Schwarzschild black hole solution.   We proceed further with the
program by analytically solving  the TeVeS equations for a static
spherically symmetric and asymptotically flat system of
electromagnetic and gravity fields.   We show that one solution is
provided by the  Reissner-Nordstr\" om metric as physical metric,
the TeVeS vector field pointing in the time direction, and a TeVeS
scalar field positive everywhere (the last feature protects from
superluminal propagation of disturbances in the fields).  We work
out black hole thermodynamics in TeVeS  using the physical metric;
black hole entropy, temperature and electric potential turn out to
be identical to those in general relativity.   We find it
inconsistent to base thermodynamics on the Einstein metric. In light
of this we reconsider the Dubovsky--Sibiryakov scenario for
violating the second law of thermodynamics in theories with Lorentz
symmetry violation.
\end{abstract}

\maketitle
s
\section{Introduction}
There are significant discrepancies between the visible masses of
galaxies and clusters of galaxies and their masses as inferred from
Newtonian dynamics.  In particular,  the accelerations of stars and
gas in the outskirts of galaxies or those of galaxies in clusters
are much too large, and the disk rotation curves of spiral galaxies,
which are naively expected to drop as $r^{-1/2}$ away from galaxy
centers, tend to remain flat to the last optically or radio measured
point. These discrepancies are also manifest in the gravitational
lensing by galaxies and clusters.  It is commonly assumed that these
problems stem from the existence in the said systems of large
amounts of ``dark matter''. For instance, galaxies are assumed to be
enshrouded in roundish halos of dark matter that dominate the
gravitational fields far from the galaxy centers.

But the putative dark matter has yet to be identified or detected
directly.  Furthermore dark matter models of galaxies require much
fine tuning of the dark halo parameters to fit the data, and there
are some sharp problems outstanding such as the observationally inferred
absence of cusps in the dark matter density at galaxy centers, cusps which
are predicted by dark matter cosmogony.  Thus many have wondered if
dark matter is the whole story.  An alternative, if less orthodox,
approach is formalized in the modified Newtonian dynamics (MOND)
paradigm~\cite{Milgrom1983}, which proposes that Newtonian gravity
progressively fails as accelerations drop below a characteristic
scale $a_{0}\simeq 10^{-10}\textrm{m}/\textrm{s}^2$  which is typical of
galaxy outskirts.   MOND assumes that for accelerations of order $ a_0$ or well below it, the Newtonian relation $\bm{a}=-\bm{\nabla}\Phi_N$ is
replaced by \be \tilde{\mu}\left(|\bm{a}|/a_0\right)\bm{a}=-\bm{\nabla}\Phi_N, \ee
where the function $\tilde \mu(x)$ smoothly interpolates between
$\tilde\mu(x)=x$ at $x\ll 1$ and the Newtonian expectation  $\tilde
\mu(x)=1$ at $x\gg 1$. This relation with suitable standard choice
of $\mu(x)$ in the intermediate range has proved successful not only
in in rationalizing the asymptotical flatness of galaxy rotation
curves where acceleration scales are much below $a_0$, but also in
explaining detailed shapes of rotation curves in the inner parts in
terms of the directly seen mass, and in giving a precise account of the observed
Tully-Fisher law which correlates luminosity of a disk galaxy with
its asymptotic rotational velocity~\cite{Bekenstein:2006}.

The pristine MOND paradigm does not fulfill the usual conservation
laws, does not make it clear if the departure from Newtonian physics
is in the gravity or in the inertia side of the equation $\bm{F}=m\bm{a}$, and does not teach us how to handle gravitational
lensing or cosmology in the weak acceleration regimes.  All these
things are done by  TeVeS~\cite{BekPRD}, a covariant field theory of
gravity which has MOND as its low velocity, weak accelerations
limit, while its nonrelativistic strong acceleration limit is
Newtonian and its  relativistic limit is general relativity (GR).
TeVeS sports two metrics, the ``physical'' metric on which all
matter fields propagate, and the Einstein metric which interacts
with the additional fields in the theory: a timelike dynamical
vector field, $u$, a scalar field, $\phi$, and a nondynamical auxiliary scalar
field $\sigma$. The theory also involves a free function
$\mathcal{F}$, a length scale $\ell$, and two positive dimensionless
constants $k$ and $K$.

TeVeS is an attempt to recast MOND into a full physical theory in
which the latter's novel behavior is due to the gravitational field.
Some checks of its consistency and comparisons with hard facts have
been made.   Thus Bekenstein showed that TeVeS's weak acceleration
limit reproduces MOND, and that it also has a Newtonian limit, and
calculated its parametrized post-Newtonian coefficients $\beta$ and $\gamma$, which agree with the results of solar system
tests~\cite{BekPRD,Bek2005,giannios:103511}. Skordis et
al.~\cite{skordis:011301} and Dodelson and
Liguori~\cite{dodelson:231301} studied the evolution of homogenous
and isotropic model universes in TeVeS, and showed that it may
reproduce key features of the power spectra of the cosmic microwave
background  and the galaxy distribution.  TeVeS has also been tested
against a multitude of data on gravitational lensing (for some
references see Ref.~\onlinecite{Bekenstein:2006}).  All the above
refer principally to situations where the gravitational potential is
small on scale $c^2$.  Since neutron stars and black holes exist in
nature one must also understand strong gravity systems in the TeVeS
framework.

A beginning in the investigation of the strong gravity regime of
TeVeS has been made by Giannios~\cite{giannios:103511}.   For vacuum
spherically symmetric and static situations he showed that under a
simplifying limit (which we shall detail below), the Schwarzschild
metric \emph{qua} physical metric and a particular scalar field
distribution together constitute a black hole solution of TeVeSs.
This motivates us to look in this paper at  more complicated cases,
such as that of the charged nonrotating black hole in TeVeS.  We
find that the Reissner-Nordstr\"om (RN) metric as physical metric
and the usual electric field  together with a special configuration
of TeVeS's scalar field constitute a black hole solution in TeVeS.
Using this solution we investigate the thermodynamics of spherical
black holes in TeVeS.

In Sec.~\ref{TeVeS} we recapitulate the fundamentals of TeVeS, while
in in Sec.~\ref{BHTeVeS} we describe Giannios' results for the
nonrotating vacuum black hole.  In Sec.~\ref{RN} we go on to solve
the TeVeS equations for the case of a charged nonrotating black
hole, obtaining a physical metric which coincides with the RN metric
of GR.  Sec.~\ref{super} presents a resolution of the problem
pointed out by Giannios: the uncharged black hole solution he found
seems to permit superluminal propagation near the black hole
horizon.   Next in Sec.~\ref{BHT} we examine how the familiar
concepts of black hole thermodynamics apply to our black hole
solutions, and check their consistency using several prescriptions.
We calculate the relevant thermodynamic quantities using the
physical metric, and show that the Einstein metric is inappropriate
for discussing thermodynamics. In this light we discuss anew  the
potential thermodynamic inconsistency described by Dubovsky and
Sibiryakov for theories with broken Lorentz
symmetry~\cite{Dubovsky}.

\section{The T\lowercase{e}V\lowercase{e}S equations\label{TeVeS}}

The acronym TeVeS refers to the Tensor-Vector-Scalar content of the
theory. The tensor part pertains to the two metrics, $g_{\mu\nu}$,
dubbed the Einstein metric, on which the vector and the scalar
fields propagate, and the physical metric $\tilde g_{\mu\nu}$, on
which matter, electromagnetic fields, etc. propagate. The
physical metric is obtained from the Einstein metric through the
following relation: \be\label{grelation}
\tilde{g}_{\alpha\beta}=e^{-2\phi}g_{\alpha\beta}-2u_\alpha u_\beta
\sinh(2\phi). \ee
Thus one passes from the space of $g_{\alpha\beta}$
to that of $\tilde g_{\alpha\beta}$ by stretching spacetime along
the vector $u$ by a factor $e^{2\phi}$, and shrinking it by the same
factor orthogonally to that vector. This prescription retains MOND
phenomenology, while augmenting the gravitational lensing by
clusters and galaxies to fit observations.

The dynamics of the metrics and the fields are derivable from an
action principle. The action in TeVeS is the sum of four terms. The first  two
 are the familiar Hilbert-Einstein action and the matter
action for field variables collectively denoted $f$:
\begin{eqnarray}
\label{HEaction}
S_g&=&\frac{1}{16\pi G}\int
g^{\alpha\beta}R_{\alpha\beta}\, \sqrt{-g}\,d^4x,
\\
S_m&=&\int \mathcal{L}\left(\tilde{g}_{\mu\nu},f^\alpha,f^\alpha_{;\mu},\cdot\cdot\cdot\right)\, \sqrt{-\tilde{g}}\,d^4x.
\label{matteraction}
\end{eqnarray}
Next comes the vector field's action ($K$ is a dimensionless positive coupling constant)
\begin{eqnarray}
S_v&=&-\frac{K}{32\pi
G}\int  \Big[\left(g^{\alpha\beta}g^{\mu\nu}u_{[\alpha,\mu]}u_{[\beta,\nu]}\right)
\nonumber
\\
&&-\frac{2\lambda}{K}\left(g^{\mu\nu}u_\mu
u_\nu+1\right)\Big]\,\sqrt{-g}\,d^4x,
\label{vectoraction}
\end{eqnarray}
which includes a constraint that forces the vector field to be
timelike (and unit normalized); $\lambda$ is the corresponding
Lagrange multiplier.  The presence of a nonzero $u^\alpha$
establishes a preferred Lorentz frame, thus breaking Lorentz
symmetry.  Finally, we have the scalar's action ($k$ is a
dimensionless positive parameter while $\ell$ is a constant with the
dimensions of length, and  ${\cal F}$ a dimensionless free
function)
 \be\label{scalaraction}S_s=-\frac{1}{2 k^2 \ell^2 G}\int
\mathcal{F}\left(k
\ell^2h^{\alpha\beta}\phi_{,\,\alpha}\phi_{,\,\beta}\right)\,\sqrt{-g}\,d^4x,
\ee
Above $h^{\alpha\beta}\equiv g^{\alpha\beta}-u^\alpha u^\beta$
with $u^\alpha\equiv g^{\alpha\beta}u_\beta$. The scalar action is here
written differently than in Ref.~\onlinecite{BekPRD}; we have
eliminated the nondynamical field $\sigma$ and redefined the
function ${\mathcal F}$. The new form makes it easier to understand
the strong acceleration limit of the theory, which is especially
relevant to the present work.

Variation of the total action with respect to $g^{\alpha\beta}$
yields the TeVeS Einstein equations for $g_{\alpha\beta}$;
\be\label{metric_eq} G_{\alpha\beta}=8\pi G\left(
\tilde{T}_{\alpha\beta}+\left(1-e^{-4\phi}\right)u^\mu\tilde{T}_{\mu(\alpha}u_{\beta)}+\tau_{\alpha\beta}\right)+\theta_{\alpha\beta}
\ee The sources here are the usual matter energy-momentum tensor
$\tilde T_{\alpha\beta}$, the variational derivative of $S_m$ with
respect to $\tilde g^{\alpha\beta}$, as well as  the energy-momentum
tensors for the scalar and vector fields:
\begin{eqnarray}
\tau_{\alpha\beta}&\equiv&
\frac{\mu}{kG}\left(\phi_{,\, \alpha}\phi_{, \,\beta}-u^\mu\phi_{,
\mu}u_{(\alpha}\phi_{, \,\beta)}\right)-\frac{\mathcal{F}
 g_{\alpha\beta}}{2k^2 \ell^2 G}\,,
\\
\nonumber
\theta_{\alpha\beta}&\equiv& K\left(g^{\mu\nu}u_{[\mu,\,\alpha]}u_{[\nu,\,\beta]}-\frac{1}{4}g^{\sigma\tau}g^{\mu\nu}u_{[\sigma,\,\mu]}u_{[\tau,\,\nu]}g_{\alpha\beta}\right),
\\
&-&\lambda u_\alpha u_\beta
\end{eqnarray}
with
\be
\mu(x)\equiv\mathcal{F}'(x).
\ee
Each choice of $\mathcal{F}$ defines a
separate TeVeS theory, and $\mu(x)$ is similar in nature to the function
$\tilde{\mu}$ in MOND. In particular, $\mu(x)\simeq 1$ corresponds
to high acceleration, i.e., to the Newtonian limit.

The equations of
motion for the vector and scalar fields are, respectively,
\begin{eqnarray}
&&\left[\mu \Big(kl^2 h^{\gamma\delta} \phi_{,\,\gamma}\phi_{,\,\delta}\Big)h^{\alpha\beta}\phi_{,\,\alpha}\right]_{;\,\beta}
\nonumber
\\
&=&kG\left[g^{\alpha\beta}+\left(1+e^{-4\phi}\right)u^\alpha
u^\beta\right] \tilde{T}_{\alpha\beta}\,,
\label{scalar_eq}
\\
&&u^{[\alpha;\beta]}\;_{;\beta}+\lambda u^\alpha+\frac{8\pi}{k}\mu
u^\beta\phi_{,\,\beta}g^{\alpha\gamma}\phi_{,\,\gamma}\nonumber
\\&=&8\pi
G\left(1-e^{-4\phi}\right)g^{\alpha\mu}u^\beta\tilde{T}_{\mu\beta}\,.
\label{vector_eq}
\end{eqnarray}
Additionally, there is the normalization condition on the vector
field:
\be
\label{normalization}
u^\alpha u_\alpha=g_{\alpha\beta}\,u^\alpha
u^\beta=-1.
\ee
The lagrange multiplier $\lambda$ can be calculated from the vector
equation.

\section{Neutral spherical Black Holes\label{BHTeVeS}}

In his work on black holes, Giannios~\cite{giannios:103511}  worked
in the limit $\mu\rightarrow 1$ which also entails ${\cal F}(x)=x$.
Since we shall later work in the same limit, we shall here justify
it in more detail than he did. Near the horizon of a black hole of
mass $m$, the Newtonian acceleration amounts to $10^{23} (M_\odot/m)
a_0$.   Thus even for the most massive black holes suspected
($10^{10}m_\odot$) the accelerations are strong on scale $a_0$ out
to at least a million times the gravitational radius, i.e. well into
the asymptotically flat region which determines the metric
properties.   This means MOND effects are suppressed while the full
complexity of the TeVeS equations is still evident.  In the said
limit, and under the assumption that the vector field points in the
time direction (which has support in the more general context of
static solutions~\cite{BekPRD}), Giannios obtained an exact
spherically symmetric analytical solution to the TeVeS equations,
for metric, scalar and vector fields.

The Einstein metric is taken in  isotropic coordinates, $x^0=t, x^1=r, x^2=\theta$ and $x^3=\varphi$,
\be\label{isotropic_metric}
ds^2=g_{\alpha\beta}dx^\alpha dx^\beta=-e^\nu dt^2+e^\zeta(dr^2+r^2
d\Omega^2),
\ee
where henceforth $d\Omega^2\equiv d\theta^2+\sin^2\theta\, d\theta^2$.
Since $\nu$, $\zeta$ are functions of $r$ only, and the
vector field points in the time direction, its $r$ dependence
is fully determined by the normalization condition \Eq{normalization}
and the requirement that $u^\alpha$ be future pointing:
\be
u^{\alpha}=(e^{-\nu/2},0,0,0).
\ee
Then the relation between the
physical and fields metrics reduces to
\begin{eqnarray}
\label{gtttrans}
&&\tilde{g}_{tt}=e^{2\phi}g_{tt},
\\
\label{grrtrans}
&&\tilde{g}_{ii}=e^{-2\phi}g_{ii}.
\end{eqnarray}

Giannios first solved the TeVeS equations assuming that $K=0$,
thus decoupling the vector field from the theory, and then performed
a transformation involving $K$, which recovered the more general solution.
For $K=0$ the Einstein's $tt, rr$ and $\theta\theta$ equations are, respectively,
\begin{eqnarray}
\frac{2\zeta'}{r}+\frac{(\zeta')^2}{4}+\zeta''&=&-\frac{4\pi(\phi')^2}{k},\label{GttK0}
\\
\frac{\zeta'+\nu'}{r}+\frac{(\zeta')^2}{4}+\frac{\zeta'\nu'}{2}&=&\frac{4\pi(\phi')^2}{k},\label{GrrK0}
\\
\frac{\nu'+\zeta'}{2r}+\frac{(\nu')^2+2\zeta''+2\nu''}{4}&=&-\frac{4\pi (\phi')^2}{k}.\label{GthetaK0}
\end{eqnarray}
and the scalar equation takes the form
\be\label{scalareqK0}
\phi''+\frac{\phi'\left(r\left(\nu'+\zeta'\right)+4\right)}{2r}=0.
\ee
Since there are only three unknown functions, $\nu(r), \zeta(r)$ and
$\phi(r)$, one of the four equations is obviously superfluous.

Combining the $rr$ and $\theta\theta$ Einstein equations gives the simple
differential equation \be
\label{nuzeta}
2(\nu+\zeta)''+\frac{6(\nu+\zeta)'}{r}+((\nu+\zeta)')^2=0. \ee This
has the solution \be \label{nu+zeta}
\nu+\zeta=2\ln\left(\frac{r^2-r_c^2}{r^2}\right), \ee where the
additive integration constant has been set to zero in order to have
an asymptotically flat spacetime, namely, $\nu,\zeta\rightarrow 0$
when $r\rightarrow \infty$.

The second integration constant, $r_c$, can be evaluated by
expanding $\nu+\zeta$ above in $1/r$ and comparing with the $1/r$
expansions (with $K=0$) of the metric coefficients of the exterior
solution for a spherical mass~\cite{BekPRD,giannios:103511},
\begin{eqnarray}
e^\nu&=&1-\frac{r_g}{r}+\frac{1}{2}\frac{r_g^2}{r^2}+\cdots
\\
e^\zeta&=&1+\frac{r_g}{r}+\frac{1}{16}\left[6-\frac{2k}{\pi}\left(\frac{G
m_s}{r_g}\right)^2\right]\frac{r_g^2}{r^2}+\cdots
\end{eqnarray}
Here $m_s$ is a mass scale~\cite{BekPRD}
defined by the expansion
\be
\label{scalarexp}
\phi(r)=\phi_c-\frac{k G m_s}{4\pi r}+\cdots
\ee
for the solution of Eq.~(\ref{scalareqK0}).
For a ball of nonrelativistic fluid, $m_s$ is very close to the Newtonian
mass, and $r_g$ is a scale of length that can be linked to the
object's mass~\cite{BekPRD}. However,  the relation between $Gm_s$ and $r_g$ depends on the
system under consideration, and is different for stars and black holes.
At any rate, for $K=0$, $r_c$ is found to be
\be\label{rcK0}
r_c=\frac{r_g}{4}\sqrt{1+\frac{k}{\pi}\left(\frac{Gm_s}{r_g}\right)^2}\,.
\ee

Making the educated \textit{guess} that \be \label{guess}
\zeta'=\frac{4r_c^2}{r(r^2-r_c^2)}-\frac{r_g}{r^2-r_c^2}\,, \ee
Giannios determines $\nu$ to be \be \label{nu}
\nu=\frac{r_g}{2r_c}\ln\left(\frac{r-r_c}{r+r_c}\right). \ee The
correctness of Eqs.~(\ref{guess}) and (\ref{nu}) can be checked by
substituting them into the sum of~\Eq{GttK0} and~\Eq{GrrK0}, or the
difference of ~\Eq{GttK0} and~\Eq{GthetaK0}; both combinations are
independent of the equation pair already used.  The determination of
the Einstein metric for $K=0$ is completed by the trivial
integration of \Eq{guess}.  Finally, the scalar field is found now
by integrating \Eq{scalareqK0} and fixing the two integration
constants just as in \Eq{scalarexp}, \be \label{scalarf}
\phi(r)=\phi_c+\frac{k Gm_s}{8\pi
r_c}\ln\left(\frac{r-r_c}{r+r_c}\right). \ee

Going on to the more general case $K\neq 0$, Giannios finds that
just by replacing Eq. (\ref{rcK0}) by \be
r_c=\frac{r_g}{4}\sqrt{1+\frac{k}{\pi}\left(\frac{Gm_s}{r_g}\right)^2-\frac{K}{2}}
\ee in the above solutions for $\nu, \zeta$ and $\phi$ will produce
an exact solution of the TeVeS equations for $K\neq 0$ [equations which are the $Q=0$ case of Eqs.~(\ref{Gtt})-(\ref{scalareq}) below].

The physical metric now follows from Eqs.~(\ref{gtttrans})-(\ref{grrtrans}):
\begin{eqnarray}
\tilde{g}_{tt}&=&-\left(\frac{r-r_c}{r+r_c}\right)^a
\\
\tilde{g}_{rr}&=&\frac{(r^2-r_c^2)^2}{r^4}\left(\frac{r-r_c}{r+r_c}\right)^{-a}
\end{eqnarray}
with $a\equiv \frac{r_g}{2r_c}+\frac{k G m_s}{4\pi r_c}$. In order
for this result to represent a black hole, the candidate event
horizon $r=r_c$ must have bounded surface area, and must not be a
singular surface. The surface area is proportional to
$\tilde{g}_{rr}(r_c)$, which has a factor $(r-r_c)^{2-a}$; for this
to be bounded requires $a\leq 2$. The Ricci scalar of the above
metric is
\be R=\frac{2(a^2-4)r_c^2r^4(r-r_c)^{a-4}}{(r+r_c)^{a+4}}\,.
\ee
 We notice that $R$ will blow up as $r\rightarrow r_c$ unless
$a=2$ or $a>4$. Thus, only the value $a=2$ describes a black hole.
The definition of $a$ then gives another relation between $r_c$ and
$r_g$,
 \be r_c=\frac{r_g}{4}+\frac{k G m_s}{8\pi}, \ee
  and the
physical metric takes the final form
\begin{eqnarray}
\label{Ggtt}
\tilde{g}_{tt}&=&-\left(\frac{r-r_c}{r+r_c}\right)^2,
\\
\label{Ggrr}
\tilde{g}_{rr}&=&\left(\frac{r+r_c}{r}\right)^4,
\end{eqnarray}
which we recognize as the Schwarzschild metric in isotropic coordinates.

Unlike GR's Schwarzschild black hole, the TeVeS neutral spherical
black hole is ``dressed''  with a scalar field $\phi$, a solution of
\Eq{scalarf}.  This field does not induce a singularity at the
horizon because of the particular structure of the TeVeS equations.
However, the logarithmic divergence of $\phi$ at the horizon was a
cause of concern to Giannios. It was earlier shown~\cite{BekPRD}
that absence of superluminal propagation of the various TeVeS fields
is guaranteed only when $\phi\geq 0$.  But here $\phi$ diverges
logarithmically at $r=r_c$, and becomes already negative
sufficiently close to $r_c$ even if $\phi_c>0$.  We will show in the
next section how this apparent problem is solved.

\section{Charged spherical black holes\label{RN}}

The next natural step is to look for an electrovacuum static
spherically symmetric solution to the TeVeS equations, the analog of
the RN solution of GR. We again take $\mu\approx 1$.   Again we
assume that the vector field points in the time direction, and that
both the physical and the Einstein metrics are spherically
symmetric. These are essential simplifying assumptions which enable
us to find a specific solution to the TeVeS field equations.  Other
solutions may exist for which the vector field is endowed with a
radial component. However, to judge from the neutral case, as
analyzed by Giannios~\cite{giannios:103511}, the PPN parameter
$\beta$ of such a solution with very low charge would be in
contradiction with recent observations~\cite{Will2003} in the solar
system.  It would be odd if the PPN structure of a black hole's far
field were that different from the sun's. By contrast, still in the
neutral case, a TeVes solution with the vector field pointing in the
cosmological time direction yields PPN parameters identical to those
of GR~\cite{giannios:103511}.

We continue to work in isotropic coordinates, for which the
transition between physical and Einstein metrics is simplest: as
seen earlier, in view of the the normalization condition
(\ref{normalization}), the transformation (\ref{grelation})  is
equivalent to \be \tilde{g}_{\alpha\beta}=
\begin{cases} e^{-2\phi}g_{ii}, & \text{$i=r$,$\theta$,$\phi$}
\\
e^{2\phi}g_{tt}\ .
\end{cases}
\ee The Einstein metric again takes the form (\ref{isotropic_metric}),
and the physical metric will have similar form, namely
\be\label{physical_metric} d\tilde
s^2=\tilde{g}_{\alpha\beta}dx^\alpha dx^\beta=-e^{\tilde{\nu}}
dt^2+e^{\tilde{\zeta}}(dr^2+r^2 d\Omega^2), \ee
with the following
relation among $\tilde{\nu}$, $\tilde{\zeta}$, $\nu$ and $\zeta$:
\begin{eqnarray}
\zeta(r)=\tilde{\zeta}(r)+2\phi(r),&&
\label{zetarel}
\\
\nu(r)=\tilde{\nu}(r)-2\phi(r).&&\label{nurel}
\end{eqnarray}

Now, the cosmological value of $\phi$ should be nonzero in our evolving universe:
$\phi(r\rightarrow\infty)=\phi_c$.   Thus  the requirement that the Einstein metric be asymptotically Minkowski (both
$\zeta$ and $\nu$ vanish as $r\rightarrow\infty$),  needed to maintain consistency
with previous work~\cite{BekPRD},  introduces a factor $\pm
2\phi_c$ in the physical metric coefficients,
\begin{eqnarray}
 \tilde{\zeta}(r\rightarrow\infty)&=&-2\phi_c,
\\
\tilde{\nu}(r\rightarrow\infty)&=&2\phi_c.
\end{eqnarray}
This is equivalent to a rescaling of the coordinates which depends on cosmological epoch, and will have to
be taken into account when considering physical quantities in the
framework of TeVeS.

The energy-momentum tensor  no longer vanishes; it
is given by
\be\label{electrotensor}
\tilde{T}_{\alpha\beta}=\frac{1}{4\pi}\left(\tilde{F}_{\alpha\rho}\tilde{F}_\beta\;^\rho-\frac{1}{4}\tilde{g}_{\alpha\beta}\tilde{F}_{\rho\sigma}\tilde{F}^{\rho\sigma}\right),
\ee
with $\tilde{F}_{\alpha\beta}$, the electromagnetic field tensor (not its dual),
obtained by solving Maxwell's equations in vacuum written wholly with the metric $\tilde g_{\alpha\beta}$, namely,
\be
\tilde{\nabla}_{\beta}\tilde{F}^{\alpha\beta}=(-\tilde g)^{-1/2}\partial_\beta\Big[(-\tilde g)^{1/2}\tilde g^{\alpha\mu} \tilde g^{\beta\nu} \tilde F_{\mu\nu} \Big]=0.
\ee
In the isotropic metric \Eq{physical_metric}, and with the assumption of
spherical symmetry and absence of magnetic monopoles, the only nonvanishing component of the electromagnetic field tensor is
\be\label{Fmunu}
\tilde{F}_{rt}=\frac{Q}{r^2}e^{\frac{1}{2}\left(\tilde{\nu}(r)-\tilde{\zeta}(r)\right)}=\frac{Q}{r^2}e^{\frac{1}{2}\left(\nu(r)-\zeta(r)\right)+2\phi(r)}.
\ee
The constant of integration $Q$  will be shown in Sec.~\ref{RN} to coincide with the physical electric charge of the black hole.

Since we assumed the vector field to point in the time direction, then
as in the vacuum case, the normalization condition
(\ref{normalization}) determines its functional dependence:
\be\label{vector}u^{\alpha}=(e^{-\nu/2},0,0,0).\ee
It follows that the spatial
components of the vector equation (\ref{vector_eq}) are identically
satisfied, while its temporal component serves to determine the Lagrange
multiplier $\lambda$ to be substituted in the Einstein equations:
\be\label{lambda}
\lambda=-\frac {K \left(r\nu'\zeta'+2r\nu''+4\nu'\right)}{4re^{\zeta
}}+\frac{GQ^2 e^{2\phi}(e^{4\phi}-1)}{r^4 e^{2\zeta}}\,.
\ee

We now turn to the Einstein equations (\ref{metric_eq}), and the
scalar equation (\ref{scalar_eq}).  Upon substitution of the Lagrange
multiplier (\ref{lambda}) and the electromagnetic field tensor
(\ref{Fmunu}), the $tt, rr$ and $\theta\theta$ equations become, respectively,
\begin{eqnarray}
\frac{2\zeta'}{r}+\frac{(\zeta')^2}{4}&+&\zeta''+\frac{K\left(8\nu'+2r\nu'\zeta'+r(\nu')^2+4r\nu''\right)}{8r}\nonumber\\
&=&-\frac{4\pi(\phi')^2}{k}-\frac{e^{-\zeta+2\phi}\,GQ^2}{r^4}\label{Gtt}
\\
\frac{\zeta'+\nu'}{r}&+&\frac{(\zeta')^2}{4}+\frac{\zeta'\nu'}{2}+\frac{K(\nu')^2}{8}\nonumber\\
&=&\frac{4\pi(\phi')^2}{k}-\frac{e^{-\zeta+2\phi}\,GQ^2}{r^4}\label{Grr}
\end{eqnarray}
\begin{eqnarray}
\frac{(\nu'+\zeta')}{2r}&+&\frac{\left((\nu')^2+2\zeta''+2\nu''\right)}{4}-\frac{K(\nu')^2}{8}\nonumber\\
&=&-\frac{4\pi
(\phi')^2}{k}+\frac{e^{-\zeta+2\phi}\,GQ^2}{r^4}\label{Gtheta}
\end{eqnarray}
The scalar equation is \be\label{scalareq}
\phi''+\frac{\left(r\left(\nu'+\zeta'\right)+4\right)\phi'}{2r}=\frac{
e^{-\zeta+2\phi}\,k GQ^2}{4\pi r^4} \ee These are four equations for
three unknowns $\zeta(r)$, $\nu(r)$ and $\phi(r)$, so one of the
equations is actually redundant. We shall use two combinations of
the three Einstein equations plus the scalar equation.

By adding the $rr$ and $\theta\theta$ equations, we again obtain
as in the vacuum case \Eq{nuzeta}.  This time we write the solution
\be\label{zeta+nu}
\zeta+\nu=2\ln\left(\frac{r^2-r_h^2}{r^2}\right).
\ee
Here one integration constant has been set so as to have an asymptotically flat spacetime, namely, $\nu,\zeta\rightarrow 0$ when $r\rightarrow \infty$.
The other constant, $r_h$, will be set by the boundary conditions on the horizon.

The remaining equations for $\nu$, $\zeta$ and $\phi$ are not
immediately solvable.  To make progress we shall \textit{assume}
that the \textit{physical} metric $\tilde g_{\alpha\beta}$ is of RN
form, solve for the scalar field in this framework, and check that
all TeVeS equations are satisfied. This will give us a pair of
charged black hole solution of TeVeS; existence of other solutions
is yet to be excluded.

In Schwarzschild coordinates $x^0=t, x^1=R, x^2=\theta$ and $x^3=\varphi$, the RN metric may be written as
\begin{eqnarray}
ds^2&=&-\left(1-R_+/R\right)\left(1-R_-/R\right)dt^2\nonumber\\
&+&\frac{dR^2}{\left(1-R_+/R\right)\left(1-R_-/R\right)}+R^2\, d\Omega^2
\end{eqnarray}
where $R_+$ and $R_-$ are the coordinates of the outer and inner
horizons, respectively. We may transform the metric to isotropic form by going over to a new radial coordinate  $r$ defined implicitly by
\be
R(r)=r+(R_+-R_-)^2/16r+(R_++R_-)/2.
\ee
 This gives
\begin{eqnarray}
d s^2=-\frac{(4r-(R_+-R_-))^2\,(4r+(R_+-R_-))^2}{(16r^2+8r(R_++R_-)+(R_+-R_-)^2)^2}dt^2\nonumber\\
+\frac{(16r^2+8r(R_++R_-)+(R_+-R_-)^2)^2}{256\,r^4}
(dr^2+r^2 d\Omega^2).\nonumber
\end{eqnarray}

We recall that in GR $R_+$ and $R_-$  satisfy the relations
$R_++R_-=2G_Nm$ and $R_+R_-=G_Nq^2$, where we write the
gravitational constant as $G_N$  to distinguish it from plain $G$,
the coupling constant in TeVeS. We shall here assume that the
\textit{physical} metric of TeVeS has the above form  while leaving
the parameters $R_+$ and $R_-$ to be determined later. However,  the
proposed metric is asymptotically Minkowskian, while as previously
mentioned, we require rather that the Einstein metric  be
asymptotically Minkowskian. This means that in the generic physical
metric \Eq{physical_metric} we must set \be
\tilde{\nu}(r)=\ln{\frac{(4r-(R_+-R_-))^2\,(4r+(R_+-R_-))^2}{(16r^2+8r(R_++R_-)+(R_+-R_-)^2)^2}}+2\phi_c,
\ee \be
\tilde{\zeta}(r)=\ln{\frac{(16r^2+8r(R_++R_-)+(R_+-R_-)^2)^2}{256\,r^4}}-2\phi_c.
\ee

To simplify these note that by Eqs.~(\ref{zetarel})-(\ref{nurel}) we have
$\zeta+\nu=\tilde{\zeta}+\tilde{\nu}$, whereupon in view of \Eq{zeta+nu},
\be
\frac{(r^2-r_h^2)^2}{r^4}=\frac{(4r-(R_+-R_-))^2\,(4r+(R_+-R_-))^2}{256\,r^4}.
\ee
We may thus relate $R_+$ and $R_-$ to the integration constant $r_h$
appearing in (\ref{zeta+nu}):
\be
\label{rh}
r_h=\frac{\scriptstyle 1}{\scriptstyle 4}(R_+-R_-).
\ee
Since $R(r=r_h)=R_+$, $r=r_h$ is the outer black hole horizon in isotropic
coordinates.
In terms of $r_h$ and $M\equiv (R_++R_-)/2$ the physical metric coefficients are
\begin{eqnarray}
\label{grrTeVeS}
e^{\tilde{\zeta}}&=&\frac{(r^2+r_h^2+Mr)^2}{r^4}\,e^{-2\phi_c},
\\
\label{gttTeVeS}
e^{\tilde{\nu}}&=&\frac{(r^2-r_h^2)^2}{(r^2+r_h^2+Mr)^2}\, e^{2\phi_c}.
\end{eqnarray}

It is useful at this point to trade the charge $Q$ for a dimensionless positive parameter $\alpha$ defined by
\be\label{GQsqr}
G\,Q^2=\alpha^2\,R_+R_-=\alpha^2(M^2-4r_h^2).
\ee
This replaces the GR relation $R_+R_-=G_Nq^2$.  The value of $\alpha$ will
be determined by the Einstein equations (\ref{Gtt})-(\ref{Grr}).

The only indeterminate function remaining now is the scalar field.  The
scalar equation (\ref{scalareq}) can be rewritten in terms of the new parameters $M$, $r_h$ and $\alpha$ as
\be
\phi''+\frac{2r\phi'}{r^2-r_h^2}-\frac{k\alpha^2(M^2-4r_h^2)e^{2\phi_c}}{4\pi(r^2+r_h^2+Mr)^2}=0.
\ee
Its general solution is
\begin{eqnarray} \phi=\phi_c+\frac{k
e^{2\phi_c}\alpha^2}{4\pi}\times
[(1+C)\ln(r+r_h)
\\
+(1-C)\ln(r-r_h)-\ln(r^2+r_h^2+Mr)]\nonumber,
\end{eqnarray}
with $\phi_c$ and $C$ integration constants, the first already
familiar.  Since we guessed the form of the metric, we need to
verify that the Einstein equations are satisfied. From the
requirement that Eq.~(\ref{Grr}) be satisfied, we obtain values for
$\alpha$ and $C$:
\begin{eqnarray}\label{alphasq}
\alpha^2&=&\frac{4\pi(2-K)e^{-2\phi_c}}{k(2-K)+8\pi}\,,
\\
\label{Cscalareq} C_\pm&=&\pm \frac{\sqrt{2k^2(2-K)+8\pi k
K}}{(2-K)k}\,. \end{eqnarray}
 \Eq{Gtt} is then satisfied identically. Since we
have already used the sum of \Eq{Grr} and \Eq{Gtheta} to get
the solution (\ref{zeta+nu}), we see that all TeVeS equations are satisfied.  Thus
the RN metric from GR with a suitable choice of parameters is the
physical metric of TeVeS spherical charged black holes.

We shall soon see that a physically acceptable solution can be had
only for $K<2$.  For such solutions the sign of the quantity under
the square root in \Eq{Cscalareq} is positive.   The two TeVeS
solutions (corresponding to the two signs of $C$) are most clearly
presented in terms of the coefficients $\delta_\pm=(k/4\pi)
\alpha^2(1+C_\pm)e^{2\phi_c}$, or \be \delta_\pm=\frac{(2-K)k
\pm\sqrt{2k^2(2-K)+8\pi k K}}{(2-K)k+8\pi}\,. \ee In view of
\Eq{vector} we finally obtain the solutions
\begin{widetext}
\begin{eqnarray}
\label{metricg}
d\tilde{s}^2&=&-\frac{(r^2-r_h^2)^2}{(r^2+r_h^2+Mr)^2}\,e^{2\phi_c}dt^2+
\frac{(r^2+r_h^2+Mr)^2}{r^4}\,e^{-2\phi_c}(dr^2+r^2 d\Omega^2),
\\
\label{scalarphi}
\phi(r)&=&\phi_c+\delta_\pm\ln(r+r_h)+\delta_\mp\ln(r-r_h)
-\frac{1}{2}(\delta_++\delta_-)\ln(r^2+r_h^2+Mr),
\\
\label{vectoru}
u^\alpha&=&\left(\frac{(r-r_h)^{\delta_\mp-1}(r+r_h)^{\delta_\pm-1}}{(r^2+r_h^2+Mr)^{(\delta_++\delta_--2)/2}},0,0,0\right).
\end{eqnarray}
\end{widetext}

\section{Resolving the superluminal paradox\label{super}}

We have found two black hole solutions for each value of $R_h$ and
$M$.  The requirement that superluminal propagation be excluded
selects one of them as physically viable.   As mentioned in
Sec.~\ref{BHTeVeS}, in a region where $\phi<0$ superluminal
propagation of the TeVeS fields is not ruled out.  This acausal
behavior would be unacceptable.  Now, since $\ln{(r-r_h)}$ in
\Eq{scalarphi} is arbitrarily large and negative near enough to the
horizon, its coefficient must be negative in order that $\phi$ have
a chance to be nonnegative everywhere.  It is easy to see that for
$K>0, k>0$, $\delta_+$ is always positive.  Thus the solution
Eqs.~(\ref{metricg})-(\ref{vectoru}) with lower signs is immediately
excluded on grounds that it permits superluminal propagation.  But
is the second solution viable in this sense?

Focusing on the solution with upper signs, we must now exclude the
parameter range $K\geq 2+8\pi/k$; the equality  here corresponds to
unbounded $\delta_-$ and $\phi$, while the inequality  leads to
$\delta_->0$ and superluminal propagation.  The range $2\leq
K<2+8\pi/k$, although palatable in this sense,  gives $\alpha^2\leq
0$.  We shall show in Sec.~\ref{BHT} that this is unphysical. For
$0<K<2$ we have $\delta_-<0$ while $\alpha^2>0$.  Thus a physically viable black hole
solution of TeVeS can exists only for $0<K<2$ (we continue to assume
that $k > 0$).  It is the solution with the lower indices in
Eqs.~(\ref{metricg})-(\ref{vectoru}).

Close enough to the horizon, $\phi$ of this solution is necessarily
positive because of the $\delta_-\ln(r-r_h)$ which is arbitrarily
large. Additionally, the asymptotic value of $\phi$ is $\phi_c$, the
cosmological value of the scalar, which may be assumed to be
positive~\cite{BekPRD}.  Hence the question of whether $\phi(r)$ is
positive in the intermediate region hinges on whether it has a
negative minimum outside the horizon, or not.

To find out we look at its derivative, \be
\phi'(r)=\frac{(M+2r_h)(r+r_h)^2\delta_-+(M-2r_h)(r-r_h)^2\delta_+}{2(r^2-r_h^2)(r^2+r_h^2+Mr)}.
\ee The numerator here is quadratic in $r$ and thus has two roots.
Now in the case $K<2$ we have $\delta_-<0$, but
$\delta_++\delta_->0$.  Then because $M>2r_h$ [see \Eq{rh} and the
following discussion], both roots are real.   Furthermore, if \be
M<2r_h\frac{\delta_+-\delta_-}{\delta_++\delta_-}=2r_h\frac{\sqrt{2k^2(2-K)+8\pi
Kk}}{(2-K)k}, \ee both roots are at  $r<0$, so  for $r>r_h$ the
field $\phi(r)$ has no minimum and must be everywhere positive.

\begin{figure}
\includegraphics[scale=0.90]{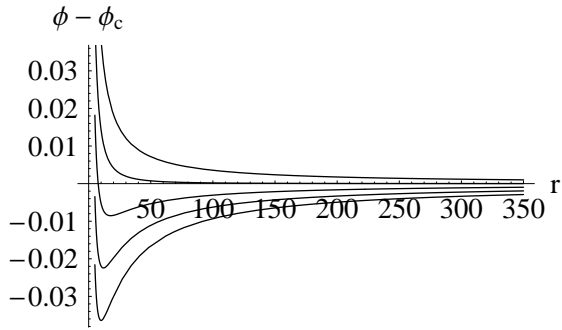}
\caption{Our solution for $\phi-\phi_c$ as a function of $r$ for several
values of $M/r_h$,  the higher $M/r_h$, the lower the curve. Both axes are in arbitrary units.\label{phidip}}
\end{figure}

Focus now on the case \be M>2r_h\frac{\sqrt{2k^2(2-K)+8\pi
Kk}}{(2-K)k}\,. \ee Now $\phi$ does have a minimum outside the
horizon.  In Fig.~\ref{phidip} we plot $\phi-\phi_c$ for several
values of $M/r_h$.  We see that unless $M/r_h$ is very large, the dip
below the axis (which grows roughly as $\ln M/r_h$) is modest
compared to unity. Hence, a modest positive $\phi_c$ (which is expected from cosmological models~\cite{BekPRD}) will be
enough to keep $\phi(r)$ positive throughout the black hole
exterior, except for black holes with exponentially large values of
$M/r_h$ for which a region of negative $\phi$ will occur near the
horizon.

In fact, for $r_h=0$, which by \Eq{rh} means that $R_+=R_-$, i.e.,
that the physical metric is extremal RN, the two solutions for
$\phi$ are identical: \be
\phi=\phi_c-\frac{(2-K)k}{(2-K)k+8\pi}\,\ln(1+\frac{M}{r}). \ee Thus
for $K<2$ and $r_h=0$, the variable part of $\phi$ is negative and
can be very large for $r\ll M$.  This is unacceptable as it permits
superluminal propagation. We may conclude that  provided $0<K<2$,
$k>0$ and $\phi_c$ somewhat above zero, the superluminality issue
raised by Giannios does not arise for the TeVeS charged black hole
solution with the lower signs in
Eqs.~(\ref{metricg})-(\ref{vectoru}).  The above conclusion does not apply to black holes near the extremally charged case.

What about Giannios' case $Q=0$ for which he found conditions
conducive to superluminal propagation (see end of Sec.~\ref{BHTeVeS}
)? The TeVeS equations (\ref{Gtt})-(\ref{scalareq}) are smooth with
$Q$, so we may take the limit $Q\rightarrow 0$ of their solutions,
Eqs.~(\ref{metricg})-(\ref{vectoru}).  In this limit according to
\Eq{GQsqr},  $M=2r_h$ while by \Eq{rh}, $r_h=R_+/4$.  Thus our
metric (\ref{metricg}) reduces exactly to Giannios'
Eqs.~(\ref{Ggtt})-(\ref{Ggrr}) with the obvious identification
$r_c=r_h$; that is we recover the fact that the physical metric is
Schwarzschild.  In the same limit  our scalar field solutions
(\ref{scalarphi}) reduce to the pair \be \phi=\phi_c+\delta_{\mp}
\ln\left(\frac{r-r_h}{r+r_h}  \right), \ee whereas Giannios obtained
only one scalar solution.  We notice that the solution with upper
sign has $\phi>0$ for all $r>r_h$ provided that we stick to the parameter ranges $0<K<2$, $k>0$ for which
$\delta_-<0$.  The solution with the lower sign has $\phi<0$
sufficiently near $r=r_h$; this is Giannios' solution, and it is
indeed excluded because it allows superluminal propagation.

To sum up, in our study of spherical static black holes in TeVeS,
we have found a viable charged black hole solution for the parameter
range $0<K<2$, $k>0$.  The limiting case $Q\rightarrow 0$ of this
is a viable neutral black hole solution.  Since black holes are seen
in nature with virtual certainty, the above results tell us that
only the range $0<K<2$, $k>0$ of TeVeS need be considered as
physical.  This range includes the values that have been explored in
the confrontation of TeVeS with
observations~\cite{BekPRD,Bekenstein:2006,skordis:011301,dodelson:231301}.

\section{Black Hole Thermodynamics
\label{BHT}}

It has been clear for long that black holes are really thermodynamic
systems characterized by temperature and entropy.  Thus a discussion
of black hole solutions in TeVeS would be incomplete without a
survey of their thermodynamic properties. However, before we can
talk about thermodynamics for the charged black hole in TeVeS, we
must first identify the physical values of  attributes of the black
hole solution.  By physical values we mean the quantities than an
asymptotically Minkowski observer would measure using instruments
made of matter, measurements which are thus referred to the physical
metric. These values need not be identical with those of quantities
naively associated with the attributes.  For example, we do not know
\emph{a priori}  that the masslike quantity $M$ and the chargelike
quantity $Q$ appearing in our solution are indeed the physical mass
and charge of the black hole. In fact, we shall see that $M$ is
related to physical mass in a nontrivial fashion.

We first note that the $G$ appearing in the TeVeS equations is not
Newton's constant, but, as shown elsewhere~\cite{BS}, is related to
it through
\be G_N=\left(\frac{(2-K)k+8\pi}{4\pi (2-K)}\right)G.
\ee
It will be useful to also write the above relation in terms of the constant $\alpha$ defined by \Eq{alphasq}:
\be\label{GNtoG} G_N=(G/\alpha^2)
e^{-2\phi_c}. \ee
Experimentally $G_N>0$; it also seems natural that
the fundamental coupling constant $G$ be positive;  hence we must
require $\alpha^2>0$. This explains why in Sec.~\ref{RN} we ruled
out the parameter range $2\leq K<2+8\pi/k$.

Next, recall that if we use the same coordinates for the Einstein and
physical metric, the transformation (\ref{grelation}) implies that
our physical metric is not asymptotically Minkowski. Thus, asymptotically, the
relation between physical distance $\tilde x$ and the corresponding spatial length-like coordinate (denoted $x$) must be
\be
\tilde x=e^{-\phi_c}x.\label{coordrel}\ee

Focus now on $M$. According to \Eq{gttTeVeS}, we may write the asymptotical expansion  for $e^{\tilde{\nu}}$ as
\be e^{\tilde{\nu}}\approx
\left(1-\frac{2M}{r}+O\left(\frac{1}{r^2}\right)\right)e^{2\phi_c}.
\ee Thus $r$ here is not physical distance $ \tilde r$, but is
related to it through Eq.~(\ref{coordrel}). Rewriting
$e^{\tilde{\nu}}$ in terms of the latter gives
\be
\label{enuassym}
e^{\tilde{\nu}}\approx \left(1-\frac{2Me^{-\phi_c}}{\tilde
r}+O\left(\frac{1}{r^2}\right)\right)e^{2\phi_c}. \ee
From the
customary linear approximation we see that $M$ and physical mass $m$
are related by \be\label{mtoMrel} Me^{-\phi_c}=G_Nm\,. \ee

Likewise, the physical charge $q$ can be easily identified by
integrating the flux of the electromagnetic field tensor through a
spherical shell at spatial infinity: \be
q=\lim_{r\rightarrow\infty}\frac{1}{4\pi}\int_{S^2}
\tilde{F}^{tr}\,e^{\tilde{\zeta}}r^2\sin\theta\, d\theta d\phi. \ee
Use of $e^{\tilde{\zeta}}$ in forming the area element guarantees
that we are calculating a physical flux: according to \Eq{zetarel}
the factor $e^{-2\phi_c}$ required by \Eq{coordrel} is supplied by
the $e^{\tilde{\zeta}}$.  Substituting $\tilde{F}^{tr}$  from
\Eq{Fmunu} gives
\begin{widetext}
\be
q=\lim_{r\rightarrow\infty}\frac{1}{4\pi}\int_{S^2} \frac{Q}{r^2}
e^{-\frac{1}{2}\left(\tilde{\nu}+\tilde{\zeta}\right)}
r^2\sin\theta\, d\theta d\phi=\lim_{r\rightarrow\infty}\frac{1}{4\pi}\int_{S^2}
\frac{Q r^2}{r^2-r_h^2} \sin\theta\, d\theta
d\phi = Q.
\ee
\end{widetext}
Thus our charged black hole is
characterized by mass $m$ and charge $Q$ as measured by physical
asymptotic observers for which the metric is $\tilde g_{\alpha\beta}$.

In investigating the black hole entropy we start with the
\emph{assumption} that it is given in terms of the physical surface
area of the outer horizon $A$ by the usual formula \be \label{SBH0}
S_{BH}=\frac{A}{4\hbar G_N}. \ee  It is true that more complicated
forms are known, but they usually appear in gravity theories with
higher derivatives; TeVeS is free of these.  The proof that our
choice is correct ultimately rests on the consistency checks we
present later in this section.

Obviously \be A=4\pi r_h^2
e^{\tilde{\zeta}(r_h)}=4\pi\left(2r_h+M\right)^2e^{-2\phi_c}. \ee
From (\ref{GQsqr}) we have for the outer horizon \be\label{rhwithM}
r_h=\frac{1}{2}\sqrt{M^2-GQ^2/\alpha^2}. \ee Thus \be \label{area}
A=4\pi\left(M+\sqrt{M^2-GQ^2/\alpha^2}\right)^2e^{-2\phi_c} \ee

We now express $A$ in terms of physical mass, charge and Newton's
constant using the relations (\ref{mtoMrel}), (\ref{GNtoG}):
\be
A=4\pi\left(G_Nme^{\phi_c}+\sqrt{(G_Nme^{\phi_c})^2-G_Ne^{2\phi_c}Q^2}\right)^2e^{-2\phi_c},
\ee
so that
\be\label{SBH}
S_{BH}=\frac{\pi}{G_N
\hbar}\left(G_Nm+\sqrt{(G_Nm)^2-G_NQ^2}\right)^2.
\ee
This is identical to the familiar expression for the entropy of a
RN black hole. To it corresponds the thermodynamic
temperature $T_{BH}=(\partial S_{BH}/\partial
m)_Q^{-1}$, or
\be\label{TBH}
T_{BH}=\frac{\hbar}{2\pi}\frac{\sqrt{(G_Nm)^2-G_NQ^2}}{\left(G_Nm+\sqrt{(G_Nm)^2-G_NQ^2}\right)^2}\,.
\ee

To check the consistency of our scheme, we now also calculate the
temperature corresponding to our black hole solution using the
Euclidean path integral approach~\cite{PhysRevD.15.2752}. This
approach entails performing a Wick rotation of the time coordinate
to obtain a Euclidean metric. The path integral for the
gravitational action then becomes the partition function for a
canonical ensemble. Regularity of the new coordinate system near the
horizon requires the new time coordinate to be periodic, and the
period is related to the black hole temperature: $T=\hbar$/period.

We first define $l$, the radial proper distance from the horizon using the physical
metric (\ref{metricg}):
\begin{eqnarray}
dl&=&\frac{(r^2+r_h^2+Mr)}{r^2}\,e^{-\phi_c}dr\,, \nonumber
 \\
\Longrightarrow
l&=&\left(r-r_h^2/r+M\ln({r}/{r_h}\right))e^{-\phi_c}.
\end{eqnarray}
Consequently the physical (2-D) line element $d\tilde \sigma^2$ for fixed $\theta$ and $\phi$ following from metric (\ref{metricg}) becomes
\be
d\tilde{\sigma}^2=-\frac{(r^2-r_h^2)^2}{(r^2+r_h^2+Mr)^2}e^{2\phi_c}dt^2+dl^2.
\ee
Near the horizon, where $r\approx r_h$, we have
\be
l\approx
(2+M/r_h)e^{-\phi_c}(r-r_h)+O\left(\left(r-r_h\right)^2\right).
\ee
Substituting this into $d\tilde{\sigma}^2$ and replacing
$e^{\phi_c}dt$, the global physical time interval according to \Eq{enuassym},  by $\imath d\tau$, we obtain an expression for the
Euclidean metric near the horizon
\be
d\tilde{\sigma}_E^2=\left(\frac{2r_h
e^{\phi_c}}{(2r_h+M)^2}\right)^2l^2(d\tau)^2+dl^2.
\ee

For this metric to be regular at $l=0$ ($r=r_h$), we must regard
$\tau$ as an angular variable with period \be
2\pi\frac{(2r_h+M)^2}{2r_h e^{\phi_c}}; \ee the corresponding
temperature is thus \be T_{BH}=\frac{\hbar r_h
e^{\phi_c}}{\pi(2r_h+M)^2}\,. \ee By means of Eqs.~(\ref{rhwithM}),
(\ref{GNtoG}) and (\ref{mtoMrel}) this can be reduced to precisely
the form (\ref{TBH}).  Thus far the thermodynamic description based
on \Eq{SBH0} is consistent.

Of course, our black hole solution must exhibit an electric
potential. By thermodynamics we would expect
that~\cite{PhysRevD.7.2333} $\Phi_{BH}=-T_{BH}(\partial
S_{BH}/\partial Q)_m$.  This gives \be\label{elecpot}
\Phi_{BH}=\frac{Q}{G_N m+\sqrt{(G_Nm)^2-G_NQ^2}}\,. \ee which agrees
with the potential of the RN black hole in GR. To verify this result
we shall also calculate the electric potential by a strictly
mechanical approach using the conservation of energy. We expect the
increase in the black hole's energy due to the fall into it of a
charged particle to equal the particle's conserved (kinetic plus
electric potential) energy.

We set out from the Lagrangian for a charged particle with mass
$\mu$ and charge $e$ \be
\mathcal{L}=e^{-\phi_c}\left(-\mu\sqrt{-\tilde{g}_{\alpha\beta}\frac{dx^\alpha}{d\tau}\frac{dx^\beta}{d\tau}}+e\tilde{A}_\alpha\frac{dx^\alpha}{d\tau}\right),
\ee where $\tilde A_\alpha$ is the potential for $\tilde
F_{\alpha\beta}$. The $e^{-\phi_c}$ normalization will soon be
justified. The Lagrangian does not depend on $t$; therefore, we have
the conserved canonical momentum \be
P_t=\frac{\partial\mathcal{L}}{\partial\frac{dt}{d\tau}}=\mu
e^{-\phi_c}\tilde{g}_{tt}\frac{dt}{d\tau}+e\tilde{A}_te^{-\phi_c}.
\ee Asymptotically $\tilde{g}_{tt}\rightarrow e^{2\phi_c}$, so that
$P_t\rightarrow\mu
e^{\phi_c}\frac{dt}{d\tau}+e\tilde{A}_te^{-\phi_c}$; now since the
physical global time is given by $\tilde{t}=e^{\phi_c}t$, the first
term in $P_t$ is recognized as minus the physical value of the
particle's rest plus kinetic energy. This justifies the
normalization we selected for the Lagrangian.  We can thus identify
$-e\tilde{A}_te^{-\phi_c}$, as the physical value of the particle's
electric energy.  The physical electric potential of the black hole
is inferred from this last energy at the horizon,
$\Phi_{BH}=-\tilde{A}_t(r_h)e^{-\phi_c}$.

We calculate $\tilde{A}_t$
by integrating the electric field (\ref{Fmunu}) from infinity to
$r_h$
\begin{widetext}
\be
\tilde{A}_t(r_h)=\int_{r_h}^{\infty}\tilde{F}_{tr}dr=-\int_{r_h}^{\infty}\frac{Q
e^{\frac{1}{2}\left(\tilde{\nu}-\tilde{\zeta}\right)}}{r^2}dr=-\frac{Qe^{2\phi_c}}{2r_h+M}\,.
\ee
\end{widetext}
Substituting here the expression (\ref{rhwithM}) for $r_h$ in terms
of $M$ and $Q$,  and switching to physical mass using \Eq{mtoMrel},
we finally get \be \tilde{A}_t(r_h)=-\frac{Qe^{\phi_c}}{G_N
m+\sqrt{(G_Nm)^2-G_NQ^2}}\,. \ee But we found the physical black hole
electric potential to be $-\tilde{A}_t(r_h)e^{-\phi_c}$, so the
present method of calculation gives exactly the same result,
\Eq{elecpot},  as the thermodynamic computation.

The above calculations serve as a consistency check of the physical
values for mass and charge which we attributed to the black hole. They
also demonstrate the physical consistency of a thermodynamical
description of spherical black holes in TeVeS. In particular, they
justify our guess (\ref{SBH0}) for the form of the black hole
entropy, and verify that the first law of black hole
thermodynamics~\cite{PhysRevD.7.2333,FourLaws} holds for the TeVeS
spherical black holes.  All this is accomplished by referring all
physics to the physical metric.  However, there is one issue for
which one must consider the role of the Einstein metric.

Recently Dubovsky and Sibiryakov (DS)~\cite{Dubovsky} showed that in
a theory with Lorentz symmetry breaking via a time-dependent scalar
field, in which there is more than one maximal propagation speed, it
would be theoretically possible to construct a perpetuum mobile that
would transfer heat from a colder to a hotter region.  This would be
accomplished, via Hawking radiation, by exploiting the different
temperatures of the nested horizons corresponding to massless fields
with different propagation speeds.

DS  consider a static spherical black hole, and two fields, $\psi_1$
and $\psi_2$, which do not interact with each other except through
gravity, and propagate at different speeds $c_1$ and $c_2$.
Consequently there exist two distinct horizons, one for field
$\psi_1$ that radiates ${\grave a\ la}$ Hawking  with temperature
$T_1$, and the second for $\psi_2$ radiating at temperature $T_2$.
It is assumed that $c_2 > c_1$; the model then gives $T_2 > T_1$. DS
assume the black hole is surrounded by two nested shells, shell $A$
which interacts with $\psi_1$ but is transparent to $\psi_2$, and
shell $B$ which interacts only with $\psi_2$. Shell $A$ has
temperature $T_A$ and shell $B$ is hotter at $T_B$.  It is also
assumed that $T_A>T_1$ and $T_B<T_2$. DS make the innocuous
assumption that heat flows from higher to lower temperature, with
the heat flow increasing monotonically with temperature difference,
and vanishing only when the two temperatures are equal.  Then they
point out that heat will flow from $A$ to the black hole via quanta
of $\psi_1$ and from the black hole to $B$ via $\psi_2$ particles.
It is possible to adjust the shell temperatures so that the two
mentioned flows become equal, in which case the black hole is in
steady state. Then the only overall effect is heat flow from $A$ to
$B$, that is from cold to hot. The second law thus appears to be
violated.

TeVeS also breaks local Lorentz symmetry, albeit by a different
mechanism: it is equipped with a timelike nonvanishing
future-pointing vector field.  Further, TeVeS possesses two metrics; this feature implies different  propagation velocities for
light and for gravitational waves.  Does the second law fail in TeVeS
within some version of the DS scenario? In order to construct the
appropriate version, we evidently first need to identify the
distinct horizons associated with light and with gravitational waves.  In
TeVeS light propagates on the null cone of the physical metric
$\tilde g_{\alpha\beta}$, and it is evident immediately from
\Eq{metricg} that the horizon for light is at $r=r_h$.  By contrast,
tensor gravitational waves propagate on the null cone of the
Einstein metric $g_{\alpha\beta}$~\cite{BekPRD}.

 With $\phi$ given by the upper sign alternative of Eq.~(\ref{scalarphi}), the Einstein
metric corresponding to the physical metric (\ref{metricg}) is
\begin{eqnarray}
\label{pmetg00}
g_{tt}=-e^{\tilde{\nu}-2\phi}=-\frac{(r-r_h)^{2-2\delta_-}(r+r_h)^{2-2\delta_+}}{(r^2+r_h^2+Mr)^{(2-\delta_--\delta_+)}}\,,&&
\\
g_{rr}=e^{\tilde{\zeta}+2\phi}=\frac{(r-r_h)^{2\delta_-}(r+r_h)^{2\delta_+}}{r^4(r^2+r_h^2+Mr)^{(\delta_-+\delta_+-2)}}\,.&&
\end{eqnarray}
For this metric the horizon (for gravitational waves) could only be
at $r=r_h$, the same location as the horizon for light!  However, as
measured with respect to $g_{\alpha\beta}$  that surface's area
diverges:  since  $\delta_{-}\leq 0$ for the physical  $K, k$
region, $g_{rr}$ blows up at horizon, and since the metric is
isotropic, this alone causes the area of the surface $r=r_h$ to blow
up.  The curvature scalars calculated with $g_{\alpha\beta}$ also
diverge at $r_h$, revealing this location to be an essential
singularity of the Einstein metric, and not a horizon. Thus we are
in no position to form a well defined entropy while working in the
geometry perceived by gravitational waves. Likewise, we cannot
obtain a black hole temperature:  no entropy, no thermodynamic
temperature.
 A similar problem arises in trying to calculate the
temperature by applying the Euclidean prescription to the Einstein
metric: the $g_{tt}$ does not behave like $l^2$.

The appearance of a singularity of $g_{\alpha\beta}$ at the same
surface as the physical metric's horizon does not pose an
insurmountable problem.  It was shown by Zlosnik, Ferreira and
Starkman~\cite{ZFS} that TeVeS can be reformulated as a
vector-tensor theory, with a single metric, the physical metric.
Probably the formal failure to bring out a thermodynamics in the
Einstein metric reflects the fact that TeVeS is at the bottom a
one-metric theory, with the Einstein metric being no more than a
mathematical convenience.

The absence of a thermodynamic temperature for gravitational waves
in the TeVeS black hole background most likely means that any
Hawking-like emission of these waves is not thermal.  The attempt,
\textit{a la} DS, to identify two distinct black hole temperatures
for the same black hole, each tied to a different maximal
propagation velocity, thus fails.  However, it has been suggested to
us that the paradox can still arise as follows.  One associates with
the black hole a \textit{graviton} effective temperature $\tau$ in
lieu of, say, $T_2$ in the DS scenario.  This $\tau$ is defined as
the temperature that the shell $B$, which interacts solely with
gravitons, would have to possess in order to just balance the black
hole's emission power in gravitons (though of course without any
pretense of detailed balance).  Now suppose $\tau$ exceeds $T_1$,
the black hole's photon temperature. By suitably adjusting $T_A$ and
$T_B$ while observing the ordering $\tau>T_B>T_A>T_1$, it should be
possible to annul the overall energy gain of the black hole.  Then
we have a DS scenario where heat flows from the low temperature
$T_A$ to the higher $T_B$ without any other change taking place.  Of
course if $\tau<T_1$, we have to arrange things with
$\tau<T_B<T_A<T_1$ to get flow from $T_B$ to the higher $T_A$.  We
emphasize that with the introduction of the effective temperature
$\tau$, the propagation velocities and Lorentz symmetry violation no
longer play the dominant role they played in the original DS
argument.

In both of the above setups we seem to have a violation of the
ordinary second law of thermodynamics.  This can be avoided only if
necessarily $\tau=T_1$, when the above schemes require $T_A=T_B$ so
that no violation is possible.  Our methods in this paper are not
suitable for the study of Hawking-like radiation, so we cannot here
affirm or exclude this possibility.  But it seems a fair conjecture
that, in fact, $\tau=T_1$.

Following the DS paper Eling, Foster, et al.~\cite{Eling} suggested
a \textit{classical} mechanism for violating the second law in its
generalized form within a gravitational theory equipped with a
timelike vector field that causes Lorentz symmetry violation.  This
mechanism also relies on two maximal propagation speeds, with
$c_A>c_B$.  It is implemented in a spherically symmetric static
situation around a black hole.  The scenario envisages a particle of
type $A$ and one of type  $B$.  They both fall through the horizon
for particles of type $A$, and interact with each other in a region
still outside the horizon for type $B$ particles, where the time
Killing vector is spacelike with respect to the metric sensed by $A$
particles.  It is then possible for the interaction to make the $A$
particle acquire negative energy while the $B$ particle's energy
remains positive.  If now $A$ falls through the $B$ particle's
horizon, and $B$ escapes passing on its way out through $A$'s
horizon, the black hole's mass has been lowered in the process.
Under mild assumptions, this means the black hole entropy decreased.
But since the particle state may have been pure all along, there is
no ordinary entropy to compensate, and so the generalized second law
is violated.

We point out that in our black hole solution there is no
intermediate region still outside the second horizon where the
Killing vector already has positive norm.  Eq.~(\ref{metricg}) shows
that in the region $r<r_h$ the Killing vector is indeed spacelike
with respect to the physical metric. However, this region is not
outside the null surface [also $r=r_h$ but in Einstein metric
Eq.~(\ref{pmetg00})] that might have been construed as the horizon
for gravitons.  Thus negative energy particles can be created, but
only inside the black hole, and the scenario envisaged in
Ref.~\onlinecite{Eling} cannot be enacted.

\section{Conclusions and Summary\label{Summary}}

We have here derived  a pair of charged spherical static black hole
solutions of TeVeS that, as far as the physical metric is concerned,
resemble the RN solution of GR.   The new features are the TeVeS
vector field which points in the time direction, and the spherically
symmetric TeVeS scalar field.    We have shown that for a wide range
of TeVeS parameters, the scalar field of one solution is positive
everywhere as long as the field has a modest positive cosmological
value.  This insures that superluminal propagation does not takes
place in that solution's background. Regarding the TeVeS analogue of
Schwarzschild's black hole earlier exhibited by Giannios, we showed
that there are actually two separate solutions here too.  For one of
them the evident positivity of the scalar field precludes
superluminal propagation.  This singles out the physical solution.

An element of guesswork entered in both Giannos' and our
derivations.  He had to guess \Eq{guess}; we had to guess the RN
form of the physical metric.  Thus in both cases it is not clear if
the black hole solutions found are the unique ones.  Proof of
uniqueness in both cases is still at large.

By expressing the parameters of the charged black hole in terms of
physical attributes measurable by material Minkowski observers, we
calculated the entropy, temperature and electric potential
characterizing the black hole.  They turn out to be the same as for
GR's RN black hole.  Black hole entropy and temperature cannot be
consistently defined for the Einstein metric, which would have been
the correct framework for studying Hawking emission of gravitational
waves.  We consider in this context a modified version of the
Dubovsky--Sibiryakov~\cite{Dubovsky} scenario for bringing about a
violation of the second law of thermodynamics out of Lorentz
symmetry breaking.  This violation of the second law can be
forestalled if a conjectured equality of the effective graviton
radiation temperature and the photon Hawking temperature holds. The
scenario described by Eling, Foster, et al.~\cite{Eling} for
bringing about classical violations of the second law in theories
with Lorentz symmetry violation cannot be be implemented with our
TeVeS black hole solutions.

\begin{acknowledgments}
This research was supported by grant  694/04  of the Israel Science Foundation, established by the Israel Academy of Sciences and Humanities.
\end{acknowledgments}

\end{document}